\documentclass[journal]{IEEEtran}

\usepackage[T1]{fontenc}
\usepackage[utf8]{inputenc}
\usepackage{graphicx}
\usepackage{amsmath}
\usepackage{amssymb}
\usepackage{epstopdf}
\usepackage{cite}
\usepackage{booktabs}
\usepackage{array}
\usepackage{multirow}
\usepackage[font=normal,labelfont=bf]{caption}
\usepackage{url}
\usepackage{xcolor}

\title{Integrated Identity Protection for Semantic Communications in Embodied IoT Systems}
\author{Yu~Zhang,
 	JingQi~Liu, 
		Shuangrui~Zhao,~\IEEEmembership{Member,~IEEE,}  
	Yuanyu~Zhang,~\IEEEmembership{Member,~IEEE,}  
		Yulong~Shen,~\IEEEmembership{Senior~Member,~IEEE,} 
		and~Xiaohong~Jiang,~\IEEEmembership{Senior~Member,~IEEE}  
		\thanks{Y. Zhang, J. Liu, S. Zhao, Y. Zhang and Y. Shen are with the School of Computer Science and Technology, Xidian University, Xi'an, China (e-mail: yuzhang\_@stu.xidian.edu.cn, dreamer\_liu2023@163.com, zhaoshuangrui@xidian.edu.cn, yyuzhang@stu.xidian.edu.cn,
        ylshen@mail.xidian.edu.cn).} 
		\thanks{X. Jiang is with the School of Systems Information Science, Future University Hakodate, Hakodate, Japan (e-mail: jiang@fun.ac.jp).}
		\thanks{	Corresponding author: Shuangrui Zhao.}
}

\begin{document}

\maketitle

\begin{abstract}
	
Embodied Internet of Things (IoT) systems are promoting the adoption of task-oriented semantic communication (SemCom), in which the trustworthy task decisions increasingly require both semantic reliability and source legitimacy.
However, existing trustworthy SemCom designs mainly focus on semantic content recovery and protection, with limited attention to source identity protection.
Motivated by this limitation, this paper establishes a new information exploitation perspective for integrated SemCom-identity protection that leverages task-level error tolerance to coordinate semantic information and source identity information carried by the same waveform.
We first examine these two information resources and then analyze their integration feasibility at the observation, processing, and decision levels.
To exploit these resources for different identity protection objectives, we further develop a dual-branch identity protection mechanism with two operating modes. 
In the authentication mode, semantic recovery and physical-layer source verification are performed in parallel using the same received observation. 
In the concealment mode,  identity exposure is suppressed while semantic reconstruction quality is maintained to support anonymous transmission.  
Finally, two representative case studies on active source impersonation and passive identity inference are presented to evaluate the proposed framework.

\end{abstract}


\section{Introduction}

The development of embodied Internet of Things (IoT) systems is driving advances in intelligent control applications, including autonomous driving, industrial automation, and human-robot collaboration.
In these systems, task-oriented semantic communication (SemCom) aims to recover task-relevant information at the semantic level to support more accurate control decisions with lower transmission latency \cite{he2026task,Shi2023task}.
However, even when the semantic information is accurately recovered, an impersonating source can inject false information and mislead the receiver's decisions.
Therefore, trustworthy task execution and control decisions in embodied IoT systems require both semantic reliability and source legitimacy.

Source legitimacy can be effectively verified through upper-layer authentication mechanisms based on cryptographic keys and certificates. Although effective, these mechanisms rely on external credentials and may introduce additional communication latency and storage overhead.
Alternatively, physical-layer authentication directly verifies source legitimacy using complementary evidence inherently embedded in the received waveform, including radio-frequency (RF) features arising from hardware impairments and channel state information (CSI) reflecting propagation conditions.

Typically, information recovery from the same received waveform also relies on physical-layer signal processing. Note that synchronization, equalization, and impairment compensation can mitigate hardware impairments and channel disturbances, thereby improving information recovery but potentially altering identity-related features. As a result, processing that improves information recovery may suppress features needed for source verification, creating a conflict between the two objectives.
Addressing this conflict calls for a task-oriented SemCom design that jointly considers semantic recovery and physical identity protection. 
Since SemCom prioritizes task-relevant information recovery over exact bit recovery, its task-level error tolerance allows certain waveform variations without compromising task performance. 
These variations can then be preserved for source verification or adjusted to reduce identity exposure, providing a design space for coordinated processing under task-performance and RF constraints.

Existing studies have provided valuable insights into semantic content recovery in SemCom and source verification through physical-layer authentication.
However, these objectives are typically investigated separately: studies focused on semantic content recovery do not necessarily address source legitimacy \cite{Shi2023task,tang2026retrieval}, while source verification methods do not necessarily account for SemCom performance \cite{xie2021survey,genaipla2026,flrff2026}.
Although some works have investigated SemCom security, they still focus primarily on protecting semantic content while overlooking its source identity  protection \cite{du2023siotsecurity,guo2024semcomnet,xu2025fssc}.
Such separated approaches overlook the need to coordinate semantic recovery and identity processing on the same received waveform, as processing choices that benefit one objective may affect the other.
Moreover, the same identity features in the waveform can be exploited by unintended observers for passive identity inference, making identity exposure suppression while maintaining semantic reconstruction quality a complementary protection objective for anonymous transmission.
To date, the coordinated exploitation of these two information resources under different identity protection objectives remains insufficiently explored.

To address this gap, we establish an integrated SemCom-identity protection framework that jointly exploits task-relevant semantic information and source identity information carried by the same waveform. Based on the feasibility of integrating these two information resources, we develop a dual-branch identity protection mechanism for source verification and identity concealment under different operating modes. A comparison with related work is provided in Table~\ref{tab:related_secure_semcom}. The main contributions are summarized as follows:

\begin{itemize}

    \item 
    We  examine these two information resources and analyze their integration feasibility based on the shared observation waveform, common processing flow and decision space compatibility. 

    \item 
     We develop a dual-branch identity protection mechanism under two operating modes, where the authentication mode jointly considers semantic reliability and source legitimacy for task input admission, while the concealment mode exploits task-level error tolerance through waveform adjustments to suppress identity exposure while satisfying task performance requirements.

    \item 
    We evaluate the proposed framework through two representative case studies, demonstrating  source verification under active source impersonation and reduced identity exposure with preserved task performance under passive identity inference.

\end{itemize}

The novelty of this work lies in a task-oriented information exploitation perspective that considers task-level error tolerance when coordinating semantic and source identity information over the shared waveform under different protection objectives.
This paper does not aim to evaluate all possible implementations of semantic representations and physical identity features. Instead, within the proposed framework, this paper first discusses what information resources are involved and why their integration is feasible from a general perspective (Section \ref{section_information}), then illustrates how the two operating modes work with representative information resources (Section \ref{sec:ldb_sf_framework}) and how task performance is evaluated (Section \ref{sec:case_study_dde_sf}). 
Finally, the research insights are summarized (Section \ref{sec_conclusion}).

\section{Information Resources in SemCom}
\label{section_information}
 
\begin{figure*}[htbp]
	\vspace{-10pt}
	\centering
	\includegraphics[width=0.7\linewidth]{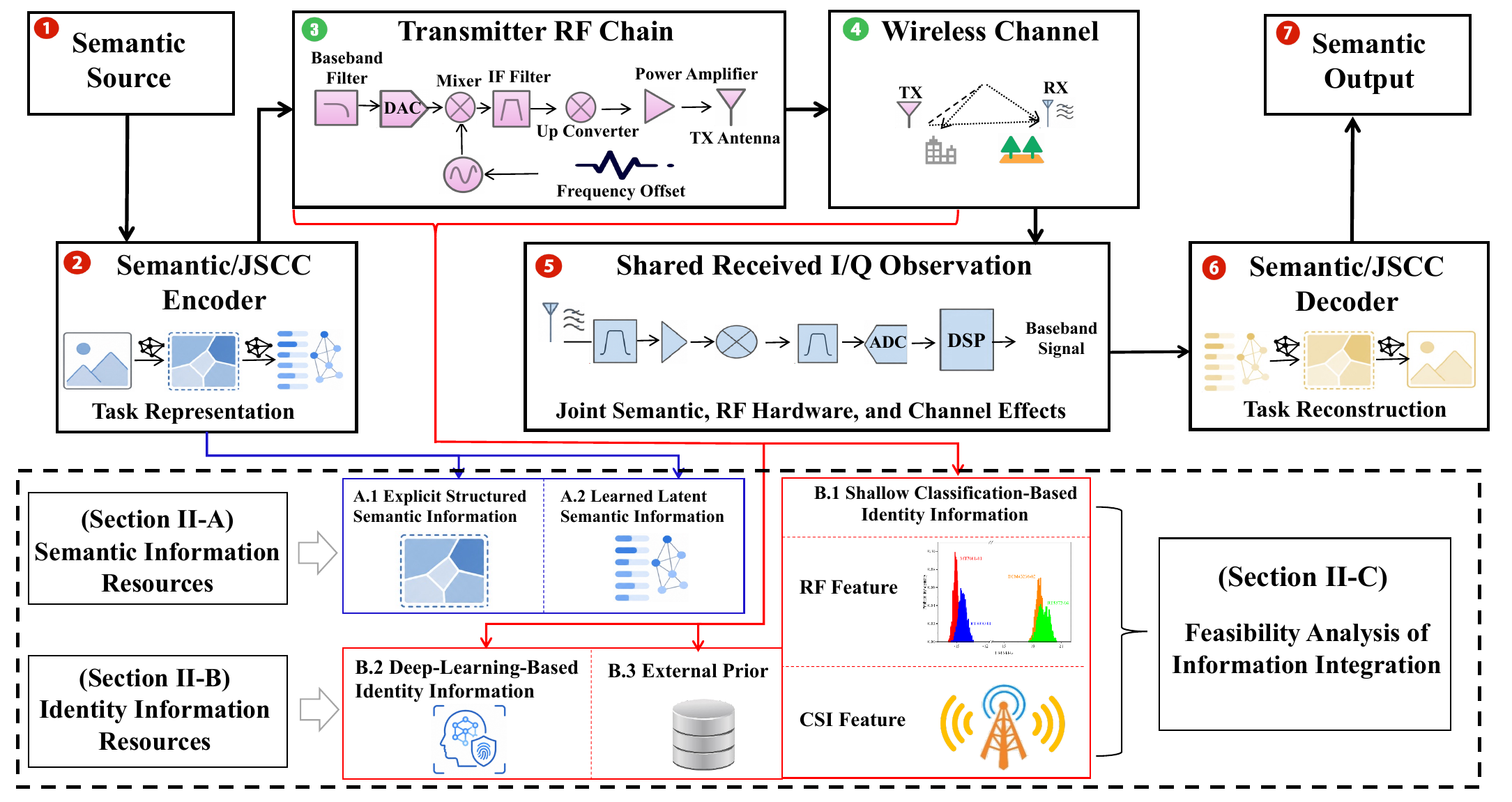}
	\vspace{-5pt}
	\caption{Lifecycle of information resources carried by SemCom signals.} 
	\label{fig:feature}
	\vspace{-15pt}
\end{figure*}

In this section, we  examine the two complementary information resources carried by SemCom signals, as illustrated in Fig.~\ref{fig:feature} and  analyze their integration feasibility to support the proposed SemCom-identity protection framework.

\subsection{Semantic Information Resources}

Semantic information resources refer to task-relevant representations extracted and encoded by the transmitter from raw data according to the communication intent. 
Depending on whether these representations are directly interpretable, they can be divided into the following two types.

 \begin{table}[t]
 	\caption{Comparison of representative studies on SemCom,
 		secure SemCom, and physical-layer authentication.}
 	\label{tab:related_secure_semcom}
 	\centering
 	\scriptsize
 	\setlength{\tabcolsep}{1.0pt}
 	\renewcommand{\arraystretch}{1.10}
 	
 	\begin{tabular*}{\columnwidth}
 		{@{\extracolsep{\fill}}lcccccc@{}}
 		\toprule
 		\textbf{Reference}
 		& \textbf{\shortstack{RF\\Auth.}}
 		& \textbf{\shortstack{CSI\\Auth.}}
 		& \textbf{\shortstack{Identity\\Concealment}}
 		& \textbf{\shortstack{GenAI\\Methods}}
 		& \textbf{\shortstack{Semantic\\Feature}}
 		& \textbf{\shortstack{Trust-Aware\\Task Decisions}} \\
 		\midrule
 		
 		\multicolumn{7}{c}{\textit{(I. SemCom)}} \\
 		
 		Shi et al. \cite{Shi2023task}
 		& $\times$ & $\times$ & $\times$
 		& $\times$ & $\checkmark$ & $\times$ \\
 		
 		Tang et al. \cite{tang2026retrieval}
 		& $\times$ & $\times$ & $\times$
 		& $\checkmark$ & $\checkmark$ & $\times$ \\
 		\midrule
 		
 		\multicolumn{7}{c}
 		{\textit{(II. Physical-Layer Authentication)}} \\
 		
 		Xie et al. \cite{xie2021survey}
 		& $\checkmark$ & $\checkmark$ & $\times$
 		& $\times$ & $\times$ & $\times$ \\
 		
 		Meng et al. \cite{genaipla2026}
 		& $\checkmark$ & $\checkmark$ & $\times$
 		& $\checkmark$ & $\checkmark$ & $\times$ \\
 		
 		Yang et al. \cite{flrff2026}
 		& $\checkmark$ & $\times$ & $\times$
 		& $\times$ & $\times$ & $\times$ \\
 		\midrule
 		
 		\multicolumn{7}{c}{\textit{(III. Secure SemCom)}} \\
 		
 		Du et al. \cite{du2023siotsecurity}
 		& $\checkmark$ & $\times$ & $\times$
 		& $\checkmark$ & $\checkmark$ & $\times$ \\
 		
 		Guo et al. \cite{guo2024semcomnet}
 		& $\checkmark$ & $\checkmark$ & $\checkmark$
 		& $\checkmark$ & $\checkmark$ & $\times$ \\
 		
 		Xu et al. \cite{xu2025fssc}
 		& $\times$ & $\times$ & $\times$
 		& $\checkmark$ & $\checkmark$ & $\times$ \\
 		\midrule
 		
 		\textbf{This work}
 		& $\checkmark$ & $\checkmark$ & $\checkmark$
 		& $\checkmark$ & $\checkmark$ & $\checkmark$ \\
 		\bottomrule
 	\end{tabular*}
 	
 	\par\vspace{2pt}
 	\begin{minipage}{\columnwidth}
 		\scriptsize
 		\textit{Note:} Auth. means authentication,
 		GenAI means generative artificial intelligence. 
 	\end{minipage}
 	\vspace{-20pt}
 \end{table}

\subsubsection{Explicit Structured Semantic Information}

Explicit structured semantic information represents task-relevant content as predefined semantic units, such as text entities, visual objects, and control instructions. 
These units can be extracted from raw data using semantic analysis methods and then encoded for transmission. At the receiver, a semantic decoder recovers the structured information to support downstream tasks.

\subsubsection{Learned Latent Semantic Information}

Learned latent semantic information is extracted using neural networks, such as convolutional neural networks (CNNs) and Transformers, and represented as compact continuous features. Although individual latent dimensions have no predefined meanings, these features jointly retain the information required for content recovery and task execution. Through deep joint source-channel coding (JSCC), the transmitter maps the latent representation into channel symbols, while the receiver reconstructs the content or directly uses the recovered representation for downstream tasks.

Despite their different forms, both types can provide task dependent reliability cues, such as output scores or reconstruction consistency. 
These cues can be used to estimate message-level semantic confidence, which can then be calibrated using task performance indicators.

\subsection{Source Identity Information Resources}

 Source identity information refers to source-related physical characteristics manifested in SemCom signals during signal generation, propagation, and reception  \cite{liang2025physical}. These characteristics can be extracted and analyzed using conventional classification or deep learning methods. 
 Such analysis can support source verification at a legitimate receiver, but may also allow a passive observer to infer source identity.

\subsubsection{Shallow Classification-Based Identity Information}

Shallow classification-based methods extract RF hardware features and channel state information (CSI) features from received signals and compare them with registered source references. 
RF hardware features mainly arise from transmitter hardware impairments, including carrier frequency offset (CFO), in-phase/quadrature (I/Q) imbalance, phase noise, and power amplifier nonlinearity. These impairments can leave distinguishable device-level features in transmitted waveforms even when devices use the same semantic encoder and communication protocol. 
In contrast, CSI mainly reflects current propagation conditions and can provide auxiliary evidence for source verification based on the expected location and link consistency \cite{genaipla2026}. 
To interpret current CSI observations, a channel knowledge map (CKM) can provide external spatial priors obtained from historical channel measurements or learning.

\subsubsection{Deep Learning-Based Identity Information}

Deep learning-based identity information does not originate from new physical sources but is learned from existing RF and CSI observations \cite{zhao2025moepls,zhao2024secdiff,mlpla2024}. 
 For example,  neural networks can extract local hardware distortion patterns and capture variations across frames and channel conditions. 
Generative models can further support data augmentation and feature reconstruction during training. 
These capabilities can improve feature robustness under varying channel and device conditions, while handling unknown transmitters requires suitable training and decision rules.

\subsection{Feasibility Analysis of Information Integration}\label{section_feasibility}

The feasibility of integrating semantic information and physical identity information can be examined through the following three factors.

\subsubsection{Shared Observation Waveform}

Semantic information and physical identity information inherently coexist in the received SemCom waveform and are observed through the same RF front end. This shared observation provides a common input for semantic recovery and identity feature extraction.

\begin{figure*}[htbp]
	\centering
	\includegraphics[width=0.82\linewidth]{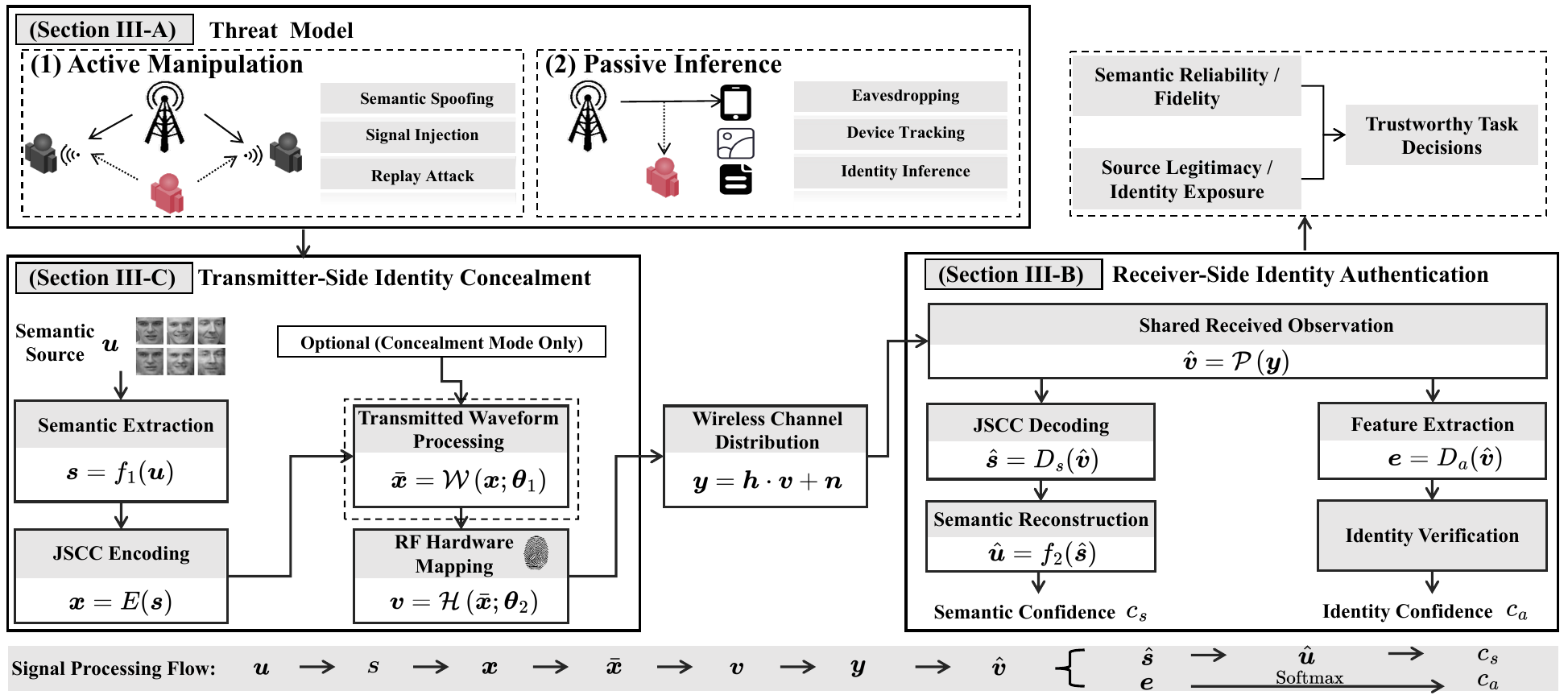}
	\vspace{-5pt}  
	\caption{Overview of the integrated SemCom-identity protection framework.
		$f_1,f_2$: semantic extraction and reconstruction;
		$E,D_s,D_a$: JSCC encoding, decoding, and identity feature extraction;
		$\mathcal W(\cdot;\boldsymbol{\theta}_1)$: optional waveform processing;
		$\mathcal H(\cdot;\boldsymbol{\theta}_2)$: RF hardware mapping;
		$\mathcal P$: receiver preprocessing;
		$\boldsymbol h,\boldsymbol n$: channel coefficient and additive noise.} 
	\label{fig:overall_framework}
	\vspace{-15pt}
\end{figure*}

\subsubsection{Common Processing Flow}

Semantic recovery and identity extraction can share early receiver processing, such as sampling, synchronization, and channel processing. However, the same operations and disturbances may affect the two processes differently, so both requirements should be considered in the joint processing design. 
Moreover, task-level error tolerance in SemCom permits certain waveform variations without compromising downstream task performance. 
Such variations may also affect identity-related features, providing a potential design space for integrating semantic recovery and physical identity protection.

\subsubsection{Compatible Decision Space}

A compatible decision space allows the semantic and identity branches to represent their outputs along two common dimensions, with their interpretation depending on the operating mode. 
In the authentication mode, semantic reliability and source legitimacy are jointly considered for task input admission. In the concealment mode, semantic fidelity and identity exposure are jointly assessed to evaluate concealment performance.

\section{Integrated SemCom-Identity Protection Framework }\label{sec:ldb_sf_framework}

In this section, we develop an integrated SemCom-identity protection framework and provide its implementation using representative information resources under different operating modes, as illustrated in Fig.~\ref{fig:overall_framework}.
The raw input $\boldsymbol u$ is mapped into semantic representations $\boldsymbol s$ and JSCC codewords $\boldsymbol x$, which undergo optional waveform processing and RF hardware mapping to generate the transmitted signal $\boldsymbol v$.
After propagation through the wireless channel $\boldsymbol h$, the receiver obtains the received signal $\boldsymbol y$ and processes it according to signal processing mechanisms. 
Accordingly, the two branches produce the recovered semantic content $\hat{\boldsymbol u}$ and identity features $\boldsymbol e$, together with their respective confidence  $c_s$ and $c_a$.
In the following, we first introduce the threat model and then describe the receiver-side authentication and transmitter-side concealment modes. In both modes, semantic and identity outputs are jointly interpreted to support task decisions based on semantic reliability and source legitimacy for authentication, or semantic fidelity and identity exposure for concealment.

\subsection{Threat Model}
\label{sec:threat_model_mapping}

In this work, we consider two representative threats to SemCom in embodied IoT systems: active source impersonation and passive identity inference.

\begin{itemize}
	\item \textbf{Active source impersonation:}
	An active attacker attempts to impersonate a legitimate source by injecting spoofed SemCom signals or replaying previously captured legitimate transmissions.
	Since such signals may remain semantically valid, the recovered semantic content can still mislead task decisions.
	
	\item \textbf{Passive identity inference:} 
	A passive eavesdropper exploits RF features from observed SemCom signals to infer source identities.
	Repeated observations may further enable cross-session linking, device tracking, and behavior profiling, potentially facilitating subsequent source impersonation.
	
\end{itemize}

These threats motivate the following two operating modes with different identity protection objectives: source legitimacy verification in the authentication mode and RF identity exposure suppression in the concealment mode.

\subsection{Identity Authentication for Trustworthy Admission Tasks }\label{sec_Authentication}

\begin{figure}[htbp]  
	\vspace{-10pt}
	\centering
	\includegraphics[width=1\linewidth]{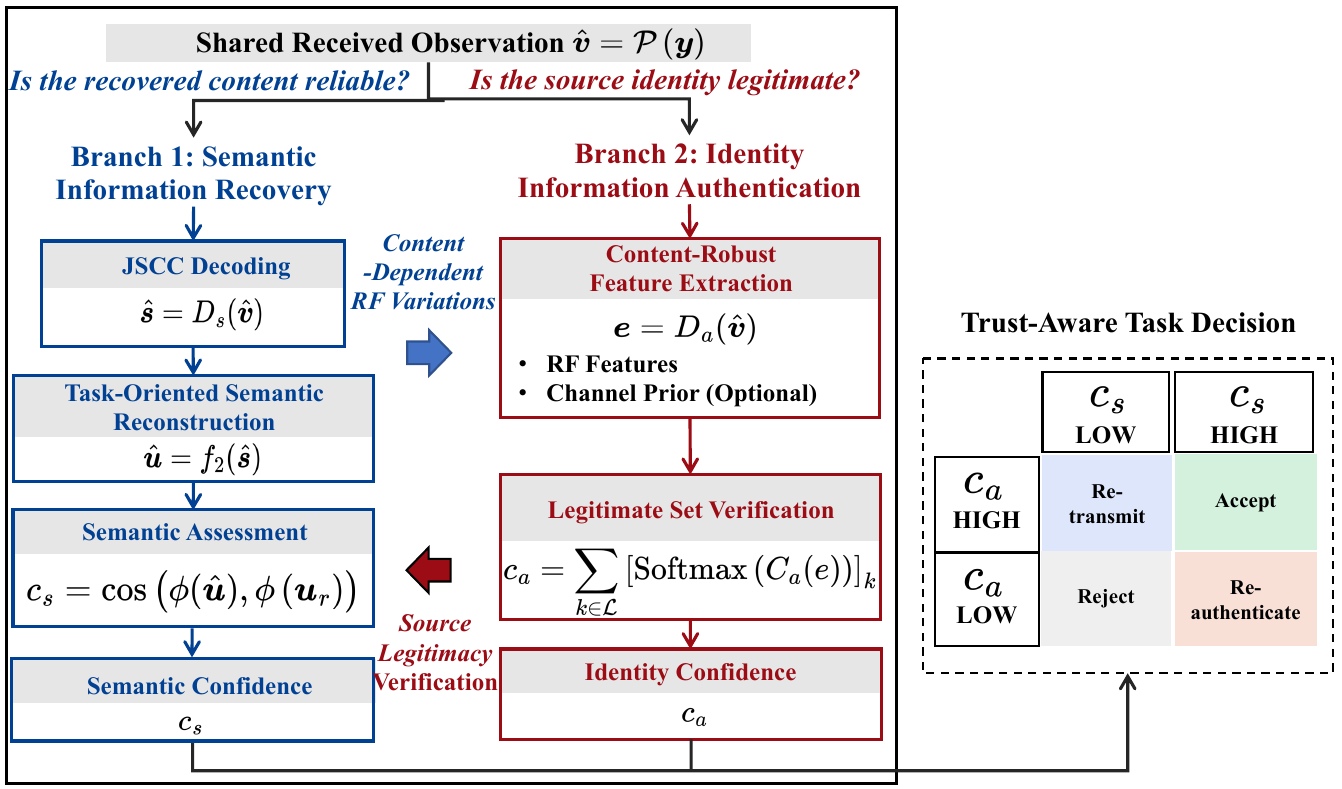}
	\vspace{-15pt}
	\caption{
		Receiver-side identity authentication mechanism. 
		$c_s$ uses cosine similarity, with $\phi(\cdot)$ extracting features from $\hat{u}$ and reference $u_r$. 
		$c_a$ sums identity classifier $C_a$'s softmax probabilities over legitimate transmitter classes $\mathcal{L}$, $k$ indexes transmitter classes. }
	\label{fig:rx_authentication}
	\vspace{-10pt}
\end{figure}

To counter active source impersonation, the receiver jointly examines semantic reliability and source legitimacy using semantic information and physical identity information from the shared received waveform. 
In this implementation, we employ signal processing techniques to reduce content-dependent variations in RF fingerprint extraction by separating the payload and preamble of the received waveform and feeding them into the semantic information recovery and identity authentication branches, respectively.
As shown in Fig.~\ref{fig:rx_authentication}, the proposed dual-branch mechanism addresses two parallel questions: 1) Is the recovered content reliable? 2) Is the source identity legitimate?

\subsubsection{Semantic Information Recovery Branch}

In embodied IoT systems, the semantic branch aims to recover task-relevant content and assess its reliability against the requirements of the current task.
To this end, the branch takes the received waveform or its processed representation after sampling, synchronization, alignment, and necessary channel processing as input.
From this input, the branch employs deep JSCC decoding followed by task-oriented semantic reconstruction to recover information required for downstream perception and decision making as discussed in Section~\ref{section_information}. 
The reconstructed information then undergoes semantic assessment against task performance requirements to obtain semantic confidence.
However, under active source impersonation, this confidence does not guarantee source legitimacy, which calls for source verification.

\subsubsection{Identity Information Authentication Branch}

This branch aims to extract source identity features from the same received waveform and verify whether the signal originates from the registered legitimate transmitter set.
To achieve this, the branch takes the processed received I/Q observation and optional channel priors as input.
It then performs content-robust feature extraction to obtain physical identity representations as described in Section~\ref{section_information}.
Legitimate-set verification evaluates the classifier probabilities assigned to registered legitimate transmitter classes and outputs an identity confidence value.

Although RF hardware fingerprints arise from relatively stable transmitter characteristics, their extraction can be affected by signal content and operating conditions.
In particular, extraction from the payload may be affected by variations in semantic content, whereas extraction from a fixed preamble can largely avoid such content effects.
However, channel disturbances may still affect fingerprint extraction even with a fixed preamble.
To improve source verification robustness, techniques such as  learning-based RF feature estimation  and representation learning across different conditions can be employed.
In the representative implementation of Fig.~\ref{fig:rx_authentication}, we focus on RF hardware fingerprints and assume that channel effects on fingerprint extraction can be sufficiently mitigated under the considered conditions.
Moreover, the shared observation waveform, common processing flow, and compatible decision space discussed in Section \ref{section_feasibility} provide a basis for jointly optimizing semantic and identity feature extraction to balance semantic recovery reliability and source verification robustness.

\subsubsection{Trust-Aware Task Decisions} 
The semantic confidence  and identity  confidence  are jointly interpreted for the current message, yielding four task admission regions.
(1) Accept region: when both confidence values are high, the system accepts the semantic information and executes the task;
(2) Re-authentication region: when semantic confidence is high but identity confidence is low, the system requests re-authentication since clear content does not imply a trustworthy source;
(3) Re-transmission region: when semantic confidence is low but identity confidence is high, the system requests re-transmission;
(4) Reject region: when both confidence values are low, the system rejects the current task.

 \subsection{Identity Concealment for Privacy Preserving Tasks} 
 
\begin{figure}[htbp]  
 	\vspace{-10pt}
 	\centering
 	\includegraphics[width=1\linewidth]{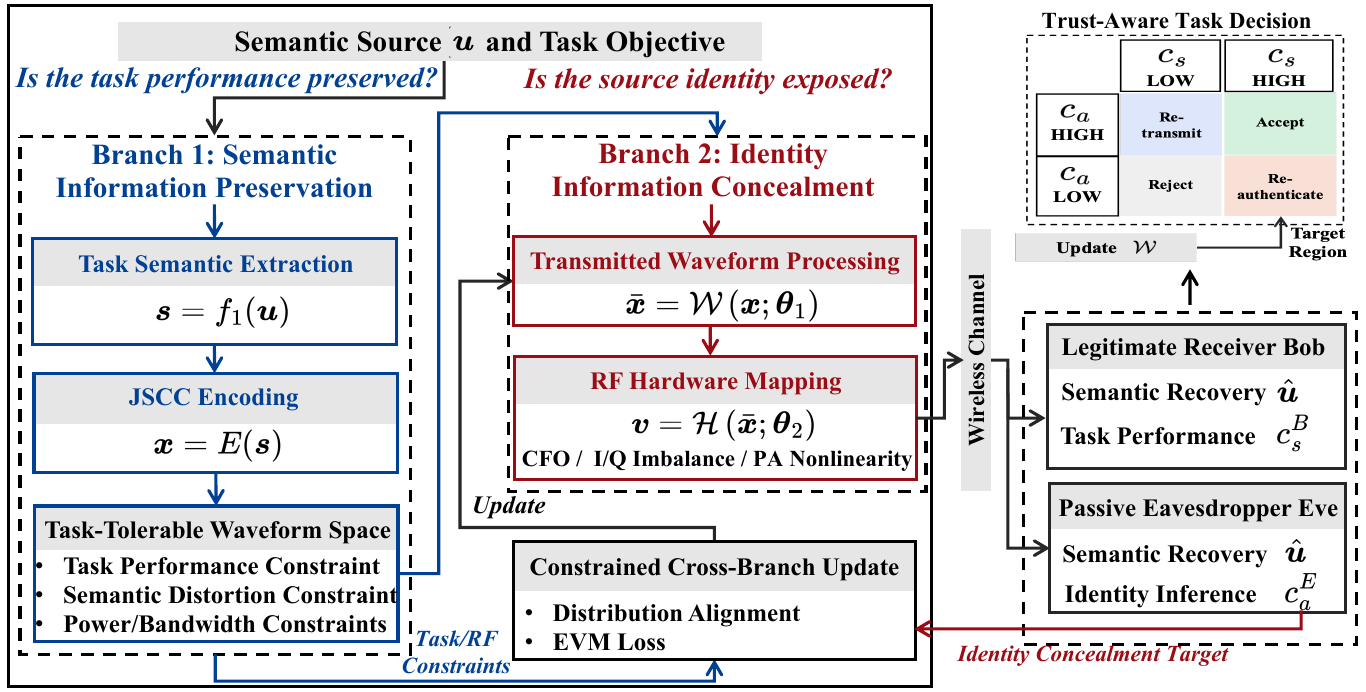}
 	\caption{Transmitter-side identity concealment mechanism.  $c_s^B$ and $c_a^E$ are semantic confidence at the legitimate receiver and identity confidence at the passive eavesdropper.}
 	\label{fig:tx_concealment}
 	\vspace{-10pt}
 \end{figure}

In embodied IoT systems, the concealment mode supports anonymous transmission that do not require disclosure of the transmitter  identity to external observers.  
Unlike receiver-side authentication, transmitter-side waveform processing in this implementation is applied to the entire transmitted signal including both the preamble and payload to suppress RF identity exposure while preserving semantic recovery quality.
  Source association relies on prior authentication or a protected session.
 To this end, we develop a dual-branch mechanism consisting of a semantic information preservation branch and an identity information concealment branch, as shown in Fig.~\ref{fig:tx_concealment}. 
 These branches coordinate task semantic requirements and physical identity protection to address two parallel questions: 1) Is the task performance preserved? 2) Is the source identity exposed?

\subsubsection{Semantic Information Preservation Branch}

The semantic branch aims to efficiently encode task-relevant information while maintaining the legitimate receiver's task performance at the required level. Given the raw data and task objectives, it employs task semantic extraction to obtain compact semantic representations, followed by deep JSCC encoding to generate the corresponding codewords.

Embodied IoT tasks vary in their task-level error tolerance. For example, minor background distortions may degrade reconstruction quality without affecting task performance, whereas changes to facial expressions may lead to incorrect decisions (see Section \ref{sec:case_study_dde_sfb}). Therefore, allowable waveform variations should be determined not only by reconstruction quality but also by task-level error tolerance.
 Accordingly, this branch establishes a task-tolerable waveform space based on the current task requirements. This space consists of candidate waveforms that satisfy task performance, semantic distortion, and power/bandwidth constraints. It thus provides task-relevant semantic representations and constraints that determine the allowable waveform variations for subsequent RF identity concealment.

\begin{figure*}[htbp]
	\centering
	\includegraphics[width=0.89 \linewidth]{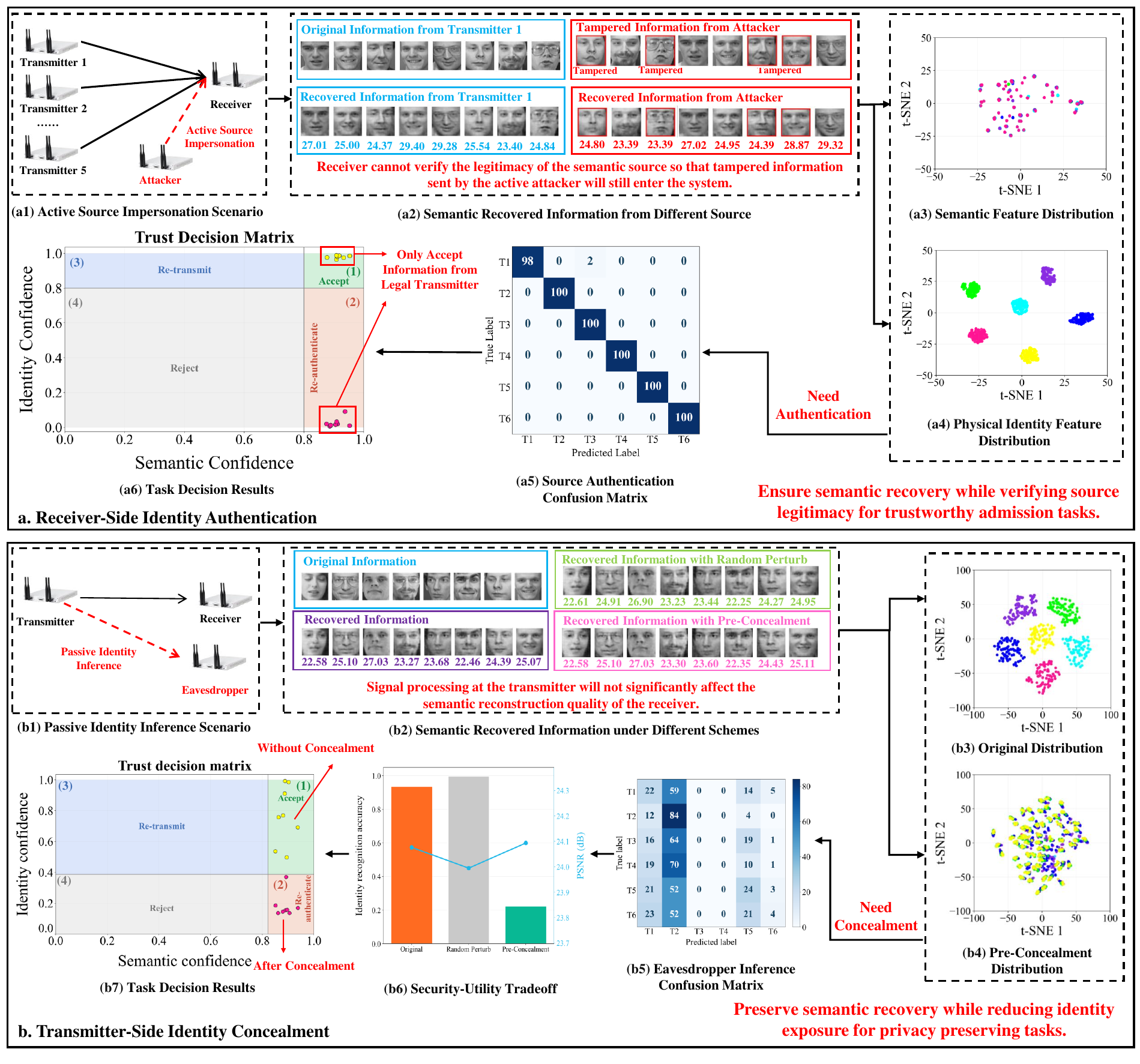}
	\vspace{-7pt}
	\caption{Case studies for trustworthy admission tasks and privacy preserving tasks.}
	\label{fig:dde_case_overview}
	\vspace{-17pt}
\end{figure*}

\subsubsection{Identity Information Concealment Branch}

The identity concealment branch aims to reduce the discriminability of RF hardware fingerprints while satisfying the legitimate receiver's task performance constraints. Although transmitter hardware characteristics are relatively stable, their observable features in the received signal can still be affected by the input waveform. 
Unlike unconstrained RF identity concealment, the proposed mechanism considers task-level error tolerance and semantic recovery requirements when determining allowable waveform adjustments, thereby reducing the ability of passive eavesdroppers to infer source identity while preserving semantic information.

Specifically, the branch uses a configurable transmitted waveform processing module to adjust the JSCC-encoded signal before transmission based on the semantic representation and task performance constraints along with pre-estimated hardware impairments and CSI. The adjusted signal then undergoes RF hardware mapping and wireless propagation, with the preceding adjustment serving as pre-compensation for the resulting hardware and channel effects. Note that directly eliminating hardware impairments or introducing arbitrary perturbations may also alter the observed identity features, but such changes may exceed the task-tolerable waveform space and degrade or even prevent semantic recovery.
After RF hardware processing and wireless propagation, the legitimate receiver evaluates semantic recovery and task performance, while a surrogate identity observer assesses source identity discriminability from the passive eavesdropper's perspective. These assessments are then fed back to guide a cross-branch update using specific utility functions (such as distribution alignment and EVM loss) to optimize the waveform processing strategy. 
Finally, the resulting waveform enables the transmitter to reduce RF identity exposure to passive eavesdroppers while satisfying the legitimate receiver's task requirements.

\subsubsection{Trust-Aware Task Decisions}

In the concealment mode, downstream task performance and RF identity exposure are evaluated at the legitimate receiver and passive eavesdropper, respectively. 
Based on these two measures, the joint assessment space is divided into four regions. The target region corresponds to satisfactory task performance at the legitimate receiver and low identity exposure at the eavesdropper.

\section{Case Studies}
\label{sec:case_study_dde_sf}

In this section, we present two representative case studies to evaluate the feasibility and effectiveness of the proposed SemCom-identity protection framework. 
All simulations were implemented in Python 3.12.6 using PyTorch 2.6.0 with CUDA 12.4 on a Windows 11 workstation equipped with an Intel Core i5-12490F CPU, an NVIDIA GeForce RTX 4060 GPU with 32 GB RAM.

\subsection{Receiver-Side Identity Authentication}
\label{sec:case_study_dde_sfa}

We evaluate receiver-side identity authentication in the active source impersonation scenario in Fig.~\ref{fig:dde_case_overview} (a1), where five legitimate transmitters (T1–T5) and an attacker (T6) transmit images to a legitimate receiver at a signal-to-noise ratio (SNR) of 20 dB\footnote{Source code: \url{https://github.com/Lab907/Identity-Sem-EmbodiedIoT/tree/main/fig5a} (accessed Nov. 5, 2026).}. 
The attacker impersonates T1 and employs region-of-interest (ROI)-based facial expression manipulation to alter the transmitted images.

Fig.~\ref{fig:dde_case_overview} (a2) shows recognizable reconstructions from both legitimate and attacker transmissions. For the eight displayed images from each source, the mean peak signal-to-noise ratios (PSNRs) are 26.11 dB for T1 and 25.77 dB for T6, measured against their respective transmitted images. The overlapping semantic feature distributions in Fig.~\ref{fig:dde_case_overview} (a3) further indicate that reconstructed image features provide limited separation between these sources. Thus, high reconstruction fidelity alone does not establish source legitimacy. In comparison, the physical identity features in Fig.~\ref{fig:dde_case_overview} (a4) form compact, well-separated transmitter clusters. The six-class confusion matrix in Fig.~\ref{fig:dde_case_overview} (a5), evaluated on 600 held-out RF signals with 100 samples per transmitter, achieves 99.67\% classification accuracy. 

Fig.~\ref{fig:dde_case_overview} (a6) illustrates the joint decision outcomes for the 16 transmissions displayed in Fig.~\ref{fig:dde_case_overview} (a2). Semantic confidence is defined as the structural-feature cosine similarity between each reconstruction and its actual transmitted image, while identity confidence is the summed classifier probability assigned to T1–T5. With both thresholds fixed at 0.8, all 16 samples satisfy the semantic criterion. Beyond these illustrative samples, authentication on the complete test set accepts all 1,500 legitimate transmissions and rejects all 300 attacker transmissions, yielding zero observed false acceptances or false rejections. These results demonstrate the value of combining semantic fidelity assessment with physical source authentication in the evaluated known-device setting.

\subsection{Transmitter-Side Identity Concealment}
\label{sec:case_study_dde_sfb}

We evaluate transmitter-side identity concealment in the passive inference scenario illustrated in Fig.~\ref{fig:dde_case_overview} (b1). Six transmitters send semantically encoded images to a legitimate receiver, while a passive eavesdropper attempts to identify the source of each transmission by analyzing its RF features. The proposed concealment mechanism aims to suppress such identity inference while preserving image reconstruction quality at the legitimate receiver\footnote{Source code: \url{https://github.com/Lab907/Identity-Sem-EmbodiedIoT/tree/main/fig5b} (accessed Nov. 5, 2026).}. 

Fig.~\ref{fig:dde_case_overview} (b2) shows that recognizable image content is preserved at the legitimate receiver under all three schemes: original transmission without additional waveform processing, random perturbation, and the proposed pre-concealment scheme. The proposed identity concealment preserves semantic reconstruction quality with negligible degradation. Fig.~\ref{fig:dde_case_overview} (b3) and Fig.~\ref{fig:dde_case_overview} (b4) show the t-distributed stochastic neighbor embedding (t-SNE) distributions of RF identity features before and after concealment, respectively. Before concealment, different transmitters form well-separated clusters, whereas the proposed method causes substantial cluster overlap, making their identities harder to distinguish. 
This is further confirmed by Fig.~\ref{fig:dde_case_overview} (b5), where the samples from other transmitters are misclassified as T2, suggesting both reduced RF-feature discriminability and a prediction bias toward T2.
As summarized in Fig.~\ref{fig:dde_case_overview} (b6), the proposed method reduces eavesdropper's identity recognition accuracy from 93.50\% to 22.33\%, while maintaining a mean PSNR of 24.09 dB compared with 24.08 dB for the original signals. 
Fig.~\ref{fig:dde_case_overview} (b7) further illustrates the decision outcomes for the 8 test samples displayed in Fig.~\ref{fig:dde_case_overview} (b2). Using fixed semantic and identity thresholds of 0.822 and 0.387, respectively, all 8 samples move from the upper-right acceptance region before concealment to the lower-right region after concealment. Their semantic confidence values remain above the semantic confidence threshold, while the eavesdropper's identity confidence values for the true transmitters fall below the identity confidence threshold. These results demonstrate that the proposed scheme preserves semantic recovery while reducing identity exposure for privacy-preserving tasks.
 
 \section{Conclusion}\label{sec_conclusion}

This paper established an integrated SemCom-identity protection framework for embodied IoT systems that coordinates semantic information and source identity information carried by the same waveform through a dual-branch mechanism, with waveform adjustments constrained by task-level error tolerance. 
The representative case studies demonstrated that source verification successfully detected impersonating sources, while identity concealment substantially reduced the evaluated eavesdropper's identity inference accuracy with negligible degradation in image reconstruction quality. 

These findings reveal that trustworthy task decisions in embodied IoT systems require not only reliable semantic recovery but also source identity protection for source verification or concealment. 
Several directions for future research are as follows. First, task-level error tolerance should be further formalized to quantify its impact on semantic recovery and identity protection. Second, authentication robustness should be further evaluated under malicious semantic injection attacks and channel variations during sensing and control in embodied IoT systems. Third, the uneven protection across transmitters calls for further investigation into improving identity concealment consistency.

\bibliographystyle{IEEEtran}
\bibliography{references}

 \section*{Biographies}
 \begingroup
 \setlength{\parindent}{0pt} 
 
 \noindent\textbf{Yu Zhang} is with Xidian University. Her research interests include physical-layer authentication.
 
 \noindent\textbf{Jingqi Liu} is with Xidian University. Her research interests include physical-layer authentication.
 
 \noindent\textbf{Shuangrui Zhao} is with Xidian University. His research interests include communication systems, physical-layer security.
 
 \noindent\textbf{Yuanyu Zhang} is with Xidian University. His research interests include radio frequency fingerprinting and GNSS security.
 
 \noindent\textbf{Yulong Shen} is with Xidian University. His research interests include  unmanned systems security and data security.
 
 \noindent\textbf{Xiaohong Jiang} is with Future University Hakodate. His research interests include wireless communications security and physical-layer authentication.
 
 \endgroup

\end{document}